\documentclass[aps,amsmath,twocolumn,amssymb,floatfix,showpacs,superscriptaddress,nofootinbib,longbibliography]{revtex4-1}
\usepackage{MnSymbol}
\usepackage{braket}
\usepackage[dvipsnames]{xcolor}
\usepackage{float}
\usepackage{subfigure}
\usepackage{tikz}
\usepackage{verbatim}
\usepackage[colorlinks=true,linktoc=page,linkcolor=violet,citecolor=magenta,urlcolor=purple]{hyperref}

\mathchardef\mhyphen="2D % Define a "math hyphen"

\newcommand{\ie}{{ i.e.,\,\,}}
\newcommand{\eg}{{e.g.,~}}

\newcommand\bea{\begin{eqnarray}}
\newcommand\eea{\end{eqnarray}}
\newcommand\beq{\begin{equation}}  
\newcommand\eeq{\end{equation}}

\usepackage[normalem]{ulem}
\definecolor{lime}{HTML}{A6CE39}
\usepackage{sidecap,tikz}
\DeclareRobustCommand{\orcidicon}{\hspace{-1.0mm}
	\begin{tikzpicture}
		\draw[lime, fill=lime] (0.0,0.0) 
		circle [radius=0.15] 
		node[white] {{\fontfamily{qag}\selectfont \tiny \,ID}};
		\draw[white, fill=white] (-0.0525,0.095) 
		circle [radius=0.007];
	\end{tikzpicture}
	\hspace{-3.0mm}
}
\foreach \x in {A, ..., Z}{\expandafter\xdef\csname orcid\x\endcsname{\noexpand\href{https://orcid.org/\csname orcidauthor\x\endcsname}{\noexpand\orcidicon}}
}

\AtBeginDocument{%
	\newwrite\bibnotes
	\def\bibnotesext{Notes.bib}
	\immediate\openout\bibnotes=\jobname\bibnotesext
	\immediate\write\bibnotes{@CONTROL{REVTEX41Control}}
	\immediate\write\bibnotes{@CONTROL{%
			apsrev41Control,author="08",editor="1",pages="1",title="1",year="1"}}
	\if@filesw
	\immediate\write\@auxout{\string\citation{apsrev41Control}}%
	\fi
}%

\begin{document}
%=============START of MAIN PAPER===============

\title{Thermoelectric properties of magic angle twisted bilayer graphene-superconductor hetero-junction: effect of valley polarization and trigonal warping}

\author{Kamalesh Bera\orcidA{}}
\email{kamalesh.bera@iopb.res.in}
\affiliation{Institute of Physics, Sachivalaya Marg, Bhubaneswar-751005, India}
\affiliation{Homi Bhabha National Institute, Training School Complex, Anushakti Nagar, Mumbai 400094, India}

\author{Pritam Chatterjee\orcidB{}}
\email{pritam@nt.phys.s.u-tokyo.ac.jp}
\affiliation{Institute of Physics, Sachivalaya Marg, Bhubaneswar-751005, India}
\affiliation{Homi Bhabha National Institute, Training School Complex, Anushakti Nagar, Mumbai 400094, India}
\affiliation{Department of Physics, The University of Tokyo, 7-3-1 Hongo, Bunkyo-ku, Tokyo 113-0033, Japan}

\author{Priyanka Mohan\orcidC{}}
\email{priya11198@gmail.com}
\affiliation{Dolat Capital, Veera Desai Industrial Estate, Andheri West, Mumbai, Maharashtra 400047, India}

\author{Arijit Saha\orcidD{}}
\email{arijit@iopb.res.in}
\affiliation{Institute of Physics, Sachivalaya Marg, Bhubaneswar-751005, India}
\affiliation{Homi Bhabha National Institute, Training School Complex, Anushakti Nagar, Mumbai 400094, India}

%--------------------------------------------------------
%--------------------------------------------------------
\begin{abstract}
	
We theoretically investigate the thermoelectric properties (electronic contribution) of a normal-superconductor (NS) hybrid junction, where the normal region consists of magic-angle twisted bilayer graphene (MATBG). The superconducting region is characterized by a common $s$-wave superconductor closely proximitized to the MATBG. We compute various thermoelectric coefficients, including thermal conductance, thermopower, and the figure of merit ($zT$), using the scattering matrix formalism. These results are further supported by calculations based on a lattice-regularized version of the effective Hamiltonian. Additionally, we explore the impact of trigonal warping and valley polarization on the thermoelectric coefficients. Notably, we find a significant variation in $zT$ as a function of these parameters, reaching values as high as 2.5. Interestingly, we observe a violation of the Wiedemann-Franz law near the charge neutrality point with the superconducting correlation, indicating that MATBG electrons behave as slow Dirac fermions in this regime. This observation is further confirmed by the damped oscillatory behavior of the thermal conductance as a function of the barrier strength when an insulating barrier is modelled at the interface of the NS junction. Beyond theoretical insights, our findings suggest new possibilities for thermoelectric applications using MATBG based NS junctions.
\end{abstract}
%--------------------------------------------------------
%--------------------------------------------------------

\maketitle
%======================================================
\section{Introduction}
%======================================================

%-------------- motivation --------------%
%Brief review of MATBG:
In recent years, magic-angle twisted bilayer graphene (MATBG) has emerged as a highly tunable material platform in quantum condensed matter domain. This van-der Walls material comes into existence in presence of a small rotational misalignment between the two graphene layers in a bilayer configuration~\cite{Li2010-vanHoveSingularity, Santos-Peres-tBLG,Shallcross-tBLG}, leading to the appearance of flat bands~\cite{MacDonald-tBLG,Moon-tBLG,Koshino-tBLG}. The discovery of unconventional superconductivity and correlated insulating phases~\cite{Cao2018-corr_insulator,Cao2018-unconv_sc} in the flat band of MATBG has brought an enormous amount of research attention in both theoretical~\cite{MIT-tbg_Liang_Fu,MI+SC-tBLG,All_Magic_Angle-tBLG,Origin_of_Magic_Angle-tBLG,CIS-tBLG,Magnetism-tBLG,Heavy_fermion-tBLG,Ferro_tBLG, KL_SC_tBLG,Kondo-tBLG-SDS,HOTS-tBLG, Arijit_kundu_tBLG,Mohan2021, bera2024} and experimental studies~\cite{Wu2021-chernIns-expt,Nuckolls2020-tBLG_xpt,Oh2021-tBLG_xpt,SOC-tBLG1,Adak2022-tDBLG,Sinha2022} of Moir\'e materials. In recent times, Josephson junctions have been fabricated using MATBG~\cite{JJ1_xpt,JJ2_xpt}. By applying local gating, a non-superconducting region is generated within the superconducting MATBG~\cite{JJ1_xpt,JJ2_xpt,JJ3_xpt,JJ4_xpt}. Such development prompts 
one to consider a heterojunction based on MATBG and theoretically investigate superconducting diode effect~\cite{JJ4_xpt} etc. %thermoelectric properties.
%\vspace{1cm}

%Review on Junctions of Graphene and Reason for studying the thermoelectric properties of MATBG

Graphene-based heterojunctions have been extensively studied over the past few years from different perspectives~\cite{CWJ-Beenakker-NS,KS_NIS,Moghaddam2006,Cuevas2006,Ludwig2007,Greenbaum2007,Bhattacharjee2007,Blanter2007,Maiti2007,Zhang2008,
Du2008,Yokoyama2008,Beenakker2008,Linder2008}. One intriguing 
outcome of those studies is the oscillatory behavior of tunneling conductance as a function of barrier strength. This is attributed to the relativistic nature of Dirac fermions in graphene’s low-energy quasiparticles. For the same reason, the thermal conductance in graphene and other Dirac-like systems also exhibits oscillatory behavior~\cite{thermal-graphene1,JL-NIS-damped-oscillation,thermal-silicene_GC_PAUL,Weyl-junction-thermal,Chatterjee-thermal-diode,Jakobsen2020,Beiranvand2017,Zhou2016,KARBASCHI2015,Wysokinski2013,Ren2013,Salehi2010,MOJARABIAN2011} with the variation of the barrier strength. However, this is in sharp contrast to the behavior of the same seen in normal metal-insulator-superconductor (NIS) junctions, where the thermal conductance decays as one increases the strength of the interfacial barrier.
%with the  barrier strength. 
The study of thermal conductance and thermoelectric properties of materials is not only of theoretical interest but also holds potential for device applications. Therefore, improving the figure of merit $zT = S^2 G T / \kappa$, where $S$, $G$, $\kappa$, and $T$ 
denote the thermopower, electrical conductance, thermal conductance, and temperature, respectively) remains a challenge in materials science. The recently engineered MATBG heterojunctions could serve as a promising candidate for further investigation.
%Therefore, improving the figure of merit remains a challenge in materials science, and the recently engineered MATBG heterojunctions can be a promising candidate for further investigation.

%existing literature on thermoelectric properties of MATBG:
Very recently, the thermal conductance (considering electronic and phononic contributions both) and thermoelectric properties of twisted bilayer graphene (tBLG) has been investigated~\cite{tblg-thermopower1,tblg-thermopower2,tblg-thermopower3,tbg_phonon_thermal_conduc1}. Also, numerous reports exist on the enhancement of thermoelectric performance in a heterojunction of different materials compared to the bulk~\cite{high_thermoelectric_of_junction1, high_thermoelectric_of_junction2,high_thermoelectric_of_junction3, high_thermoelectric_of_junction4, high_thermoelectric_of_junction5,PhysRevB.110.245417}. However, the thermal transport across a hybrid structure \eg normal-superconductor (NS) junction comprised of MATBG has not been 
explored yet. 
%There is numerous reports on the enhancement of thermoelectric metrics in a hetero junction compared to the bulk material~\cite{high_thermoelectric_of_junction1, %high_thermoelectric_of_junction2,high_thermoelectric_of_junction3, high_thermoelectric_of_junction4, high_thermoelectric_of_junction5}. 
This motivates us to investigate the thermal conductance and thermoelectric properties across the hybrid junction comprised of MATBG. %and MATBG superconductor.

%-------------- main findings --------------%
%Model, a brief intro & findings
In this article, we explore the thermoelectric properties through a NS hybrid junction based on MATBG. Here, we assume that the superconductivity has been induced in MATBG via the proximity effect.
In the first part of our study, we calculate the thermoelectric coefficients based on a quasi-one-dimensional (1D) version of an effective two-dimensional (2D) model for MATBG~\cite{MIT-tbg_Liang_Fu,Koshino-tBLG,KT_Law_JDE-TBG1,KT_Law_JDE-TBG2} using Onsager relations and the scattering matrix formalism. In the latter part, we further validate our findings with a 
lattice-regularized version of the effective continuum model. We observe that the results obtained from the lattice model align well with those proposed from the continuum model, with some minor thermal fluctuations and finite-size effects in the high-temperature regime. Interestingly, we find a violation of the Wiedemann-Franz (WF) law near the charge neutrality point, indicating that MATBG electrons behave as slow Dirac fermions. We further confirm this observation through the damped oscillatory behavior of thermal conductance as a function of barrier strength in the presence of an insulating barrier inserted between the NS junction. Additionally, we observe an enhanced figure of merit $zT$ value due to the interplay between the trigonal warping of MATBG and valley polarization, suggesting that our system can serve as a promising thermoelectric candidate with potential for future device applications. 

%-------------- Structure --------------%
The remainder of this paper is organized as follows. In Sec.~\ref{Sec:II}, we introduce our model Hamiltonian and the scattering matrix formalism. In Sec.~\ref{Sec:III}, we define the various thermoelectric coefficients using the Onsager relations. In Sec.~\ref{Sec:IV}, we present our findings based on both the continuum and lattice models, with an emphasis on the impact of trigonal warping and valley polarization on the system's $zT$ value. In Sec.~\ref{Sec:V}, we examine the effect of an insulating barrier inserted at the interface of the MATBG NS junction, followed by a summary and discussion 
in Sec.~\ref{Sec:VI}.

%======================================================
\section{Model and Method}\label{Sec:II}
%======================================================
In this section, we present the model Hamiltonian for our setup and discuss the scattering matrix formalism to investigate thermal transport across the NS hybrid junction of MATBG. Additionally, we construct a real-space tight-binding model based on a lattice-regularized version of the effective Hamiltonian to validate our results.

%-----------------------------------------------------------------------------------
%-----------------------------------------------------------------------------------
\begin{figure}[t]
	\centering
	\subfigure{\includegraphics[width=0.52\textwidth]{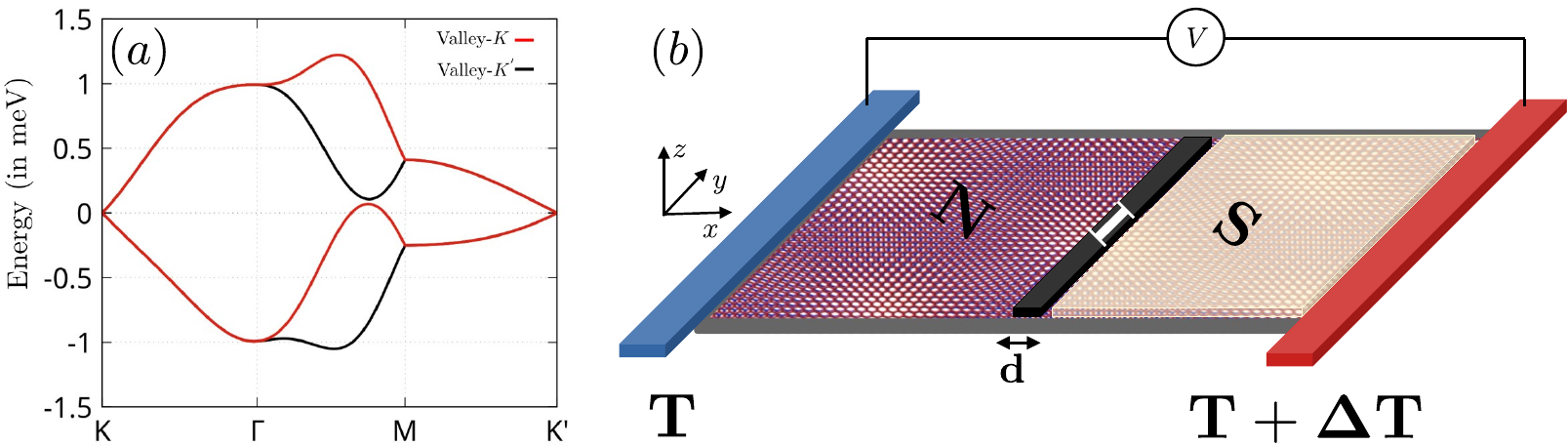}}
	\caption{In panel (a), we illustrate the band dispersion of MATBG derived from the effective two-orbital tight-binding model with the chosen parameters $t_{1} = 0.331$ meV, $t_{2} = -0.01$ meV, 
	and $\tilde{t}_{2} = 0.097$ meV. In panel (b), we depict a schematic diagram of our setup, where the left region of the hybrid geometry represents the MATBG (labelled as normal ``N"), while 
	the right region denotes the proximity induced MATBG superconductor with $s$-wave pairing (labeled as superconductor ``S"). An insulating barrier (MATBG with different Fermi energy) 
	of width $d$ (labeled as insulator ``I") is placed at the NS interface. The thermal baths are connected to the left and right sides of the hybrid structure and are depicted in blue and red, 
	maintaining temperatures $T$ and $T + \Delta T$, respectively, along with an applied voltage bias $V$.
}
 \label{fig1}
\end{figure}
%--------------------------------------------------------------------------------
%--------------------------------------------------------------------------------

%--------------------------------------------------------------------------------
\subsection{Low energy continuum model and scattering matrix formalism}
%--------------------------------------------------------------------------------
Since the flat bands in tBLG are isolated from rest of the conduction and valence bands, it is possible to construct an effective 2D tight binding model (valid only near the magic angle) maintaining the 
%which keeps the necessary symmetries intact 
necessary symmetries~\cite{MIT-tbg_Liang_Fu,Koshino-tBLG}. Here, we provide a brief outline to derive the low-energy continuum model Hamiltonian starting from the effective two-orbital 
tight-binding model of MATBG. The model Hamiltonian that represents the relevant Moir\'e bands near charge neutrality can be expressed as,

\begin{eqnarray}
	\begin{aligned}
	H_{0} &= - \sum_{n, \sigma \tau}\mu_{n} c^{\dagger}_{n\sigma\tau} c_{n\sigma\tau} + \sum_{\langle n m\rangle, \sigma\tau} t_{1}  c^{\dagger}_{n\sigma\tau} c_{m\sigma\tau} + \text{h.c.}\\
	 &+ \sum_{\langle n m\rangle^{'}, \sigma\tau} (t_{2} -i\sigma\tau \tilde{t}_{2} ) c^{\dagger}_{n\sigma\tau} c_{m\sigma\tau}  + \text{h.c.}	\ ,
	\end{aligned}
 \label{Eq1}
\end{eqnarray}
 where, $\tau = \pm1$ represents the valley index corresponding to valleys $K$ and $K^{\prime}$, respectively, while $\sigma = \pm1$ denotes the spin indices, corresponding to spin-up and spin-down, respectively. The notation $\langle n m \rangle$ represents the nearest-neighbor hopping on a hexagonal lattice with hopping amplitude $t_{1}$. Similarly, $\langle n m \rangle^{'}$ denotes the fifth-nearest-neighbor hopping element with amplitude $(t_{2} -i\sigma\tau \tilde{t}_{2} )$. The term associated with $\tilde{t}_{2}$ acts as a trigonal warping term and the symbol $\mu$ represents the chemical potential. The operator $c_{n\sigma\tau}$ ($c_{n\sigma\tau}^{\dagger}$) denotes the annihilation (creation) operator of an electron at the $p$-wave-like orbital $p_{x} + i \tau p_{y}$ at site $n$ with spin $\sigma$. In Fig.~\ref{fig1}(a), we illustrate the band structure obtained using the momentum-space version of the model Hamiltonian, with the parameters $t_1 = 0.331$ meV, $t_2 = -0.01$ meV, 
and $\tilde{t}_2 = 0.097$ meV. Note that, this nearly corresponds to the flat band structure obtained for full tBLG~\cite{MacDonald-tBLG,Moon-tBLG,Koshino-tBLG}.
%To get a more accurate description of the flat bands one needs to consider more hopping in the real space model as shown in ref.[]

We also take into account the effect of valley polarization, which arises due to electron-electron interactions as demonstrated in Ref.~\cite{KT_Law_JDE-TBG2}, in the following way,
\begin{eqnarray}
	\begin{aligned}
		H &= H_{0} + \sum_{n \sigma\tau} \Delta_{vp} c^{\dagger}_{n\sigma\tau} (\tau_{z})_{\tau\tau^{'}} c_{n\sigma\tau^{'}}\ ,	
	\end{aligned}
	\label{Eq2}
\end{eqnarray}
where, $\Delta_{vp}$ is the valley polarization order parameter with the Pauli matrix $\tau_{z}$. 

From the momentum space version of the 2D tight-binding model Hamiltonian [Eq.~(\ref{Eq2})], one can derive a $k \cdot p$ model around the $\Gamma$-point employing a Taylor series expansion. Consequently, the effective low-energy continuum model for MATBG with interaction-induced valley polarization takes the form~\cite{KT_Law_JDE-TBG1,KT_Law_JDE-TBG2},
\begin{eqnarray} 
H^{\tau}_{\rm {2D}} = -\mu + \lambda_{0} (k_{x}^{2} + k_{y}^{2})  +   \lambda_{1} \tau  (k_{x}^{3} - 3 k_{x} k_{y}^{2}) + \tau \Delta_{vp}\ .
\label{Eq3}
\end{eqnarray} 

Here, $\lambda_{0}$ and $\lambda_{1}$ denote the kinetic energy and the trigonal warping term, respectively. The symbol $\Delta_{vp}$ represents the degree of valley polarization. A non-zero value 
of $\Delta_{vp}$ breaks the degeneracy between the two valleys, thereby breaking time-reversal symmetry.   

The quasi-1D continuum version of the above model [Eq.~(\ref{Eq3})], can be expressed by setting the transverse momentum $k_y=0$. This can be written as,
\begin{eqnarray}\label{eq3}
h^{\tau} = \lambda_{0} k_{x}^{2} + \lambda_{1} \tau  k_{x}^{3} + \tau \Delta_{vp} -\mu\ .
\label{Eq4}
\end{eqnarray} 
This Hamiltonian [Eq.~(\ref{Eq4})] effectively describes the band structure near the $\Gamma$-point mentioned in Fig.~\ref{fig1}(a). 

%Context of BdG Hamiltonian, temperature dependent gap, validity of BCS-MF,
In Fig.~\ref{fig1}(b), we depict the schematic of our MATBG based NS junction in the presence of an insulating barrier inserted at the NS interface. The normal region is defined for $x \leq -d$, and the insulating barrier extends from $x = -d$ to $x = 0$. The region for $x \geq 0$ is the proximity induced superconducting region. The insulating barrier is modeled by a barrier potential of strength $V_0$, which can be realized by applying an electric field or gate voltage. A temperature gradient is established across the NS junction by employing two thermal baths attached to the left and right sides 
of the junction.

In order to linearize the Hamiltonian in Eq.~(\ref{Eq4}) near the Fermi momentum, we apply the Andreev approximation~\cite{Andreev-approximation}. Assuming the proximity induced superconducting pairing exhibits a conventional \(s\)-wave nature, the Bogoliubov-de Gennes (BdG) Hamiltonian describing the linearized system in the presence of the superconductor can be written as,
\begin{equation}
	\begin{aligned}
	H_{\tau, \alpha} = 
	\left( \begin{array}{cc}
	h_{\tau, \alpha} & \Delta_{s}(x)\\
	\Delta_{s}^{\dagger}(x) & -h_{-\tau, -\alpha}^{*} \end{array}\right)\ , 
	\end{aligned}
 \label{Eq5}
\end{equation}
where, $h_{\tau, \alpha} = - \alpha \mathrm{i} \hbar v_{\tau, \alpha}(x) \partial_{x} + \tau \Delta_{vp}(x)$. The expression for the longitudinal Fermi velocity $v_{\tau, \alpha}(x)$ is provided in Appendix \ref{AppA}. The index $\alpha$ takes the values $\pm 1$, representing forward and backward moving electrons, respectively, along the $x$ direction. The above Hamiltonian is written in terms of the Nambu basis, $[\psi_{\tau,\alpha}, \psi_{-\tau,-\alpha}^{\dagger}]^{T}$. The forms of the superconducting pair potential and valley polarization can be expressed as $\Delta_{s}(x) = \Delta_s(T) e^{i\phi} \theta(x)$ and $\Delta_{vp}(x) = \Delta_{vp} \theta(-x)$, respectively, where $\theta(x)$ is the Heaviside step function. We account for the temperature dependence of the superconducting gap $\Delta_s(T)$ using the relation $\Delta_{s}(T) = \Delta \tanh(1.74\sqrt{(T_{c} - T)/T})$, where $\Delta = \Delta_s(0)$ is the pairing gap at zero temperature and $T_c$ is the superconducting critical temperature.

We define the chemical potential in the normal and superconducting regions as $\mu = \mu_N$ and $\mu = \mu_S = \mu_N + U_0$, respectively, where $U_0$ is a constant back gate voltage in the superconducting region. Throughout our numerical calculations, we assume the doping in the superconducting region is very high (\ie $U_0 \gg \mu_N$) to ensure that the mean-field condition for superconductivity is satisfied~\cite{CWJ-Beenakker-NS}. The chemical potential in the insulating barrier region can be express as, $\mu=\mu_N-V_0$.

%Scattering matrix formalism:

 In order to obtain the possible scattering amplitudes from the NS interface, the wave functions in the normal ($\Psi_{N}$), insulating ($\Psi_{I}$), and superconducting ($\Psi_{S}$) regions 
 can be written as,
\begin{equation}
	\begin{aligned}
		\Psi_{N} &= \psi^{N}_{\overrightarrow{e}} + r_{e}  \psi^{N}_{\overleftarrow{e}} +  r_{A}  \psi^{N}_{\overleftarrow{h}}\ , \\
		\Psi_{I} &= a \psi^{I}_{\overrightarrow{e}} +  b  \psi^{I}_{\overleftarrow{e}} + c \psi^{I}_{\overrightarrow{h}} +  d  \psi^{I}_{\overleftarrow{h}}\ ,\\
		\Psi_{S} &= t_{e} \psi^{s}_{\overrightarrow{eq}} +  t_{h}  \psi^{s}_{\overleftarrow{hq}} \ .
	\end{aligned}
 \label{Eq6}
\end{equation}

Here, the symbols $r_e$ and $r_A$ represent the normal and Andreev reflection (AR) coefficients, respectively. On the other hand, $t_e$ and $t_h$ denote the tunneling amplitudes of the electron-like and hole-like quasiparticles in the superconducting region. Note that, $t_e$ and $t_h$ become vanishingly small due to decaying wave-function in the subgapped regime $E < \Delta$. The symbols $a$, $b$, $c$, and $d$ indicate the scattering amplitudes of the electrons and holes in the insulating barrier region. The notation denoted by the right and left arrows represent right- and left-moving 
particles (electrons/holes), respectively. The detailed analytical expressions for the spinors are provided in the Appendix~\ref{AppA}.

In order to evaluate the scattering coefficients, we employ the appropriate boundary condiction in terms of continuity of the wave functions at $x = -d$ and $x = 0$ [using Eq.~(\ref{Eq6})]. This leads to the following boundary conditions,
\begin{equation}
	%\begin{aligned}
		\Psi_{N} (x = -d) = \Psi_{I} (x = -d)\ , \hspace{10pt}
		\Psi_{I} (x = 0) = \Psi_{S} (x = 0)\ ,
	%\end{aligned}
	\label{Eq7}
\end{equation}
Here, continuity of the wave functions at the boundaries is sufficient to obtain the scattering coefficients as Eq.~(\ref{Eq5}) exhibits a linearized Hamiltonian.

Although it is quite cumbersome to derive an analytical expression for the scattering coefficients of a MATBG based NS junction in the presence of an insulating barrier at the interface, it is possible to obtain the analytical expressions in the absence of such a barrier. Therefore, such expressions for the normal and AR amplitudes in the absence of an insulating barrier can be expressed as, 

\begin{equation}
	\begin{aligned}
	r_{e} &= \frac{(e^{2 i \beta } - 1) \sqrt{N_{e-}} (v_{s-} + v_{n+})  (v_{n+} - v_{s+})} {\sqrt{N_{e+}} \big( A  + e^{2 i \beta}  B \big)}\ , \\
	r_{A} &= \frac{e^{i \beta }  \sqrt{N_{h+}} (v_{n-} + v_{n+})  ( v_{s-} + v_{s+})} {\sqrt{N_{e+}} \big( A  + e^{2 i \beta }  B \big)}\ ,
	\end{aligned}
 \label{Eq8}
\end{equation}

%The normalization factors,
%\begin{align*}
%	N_{e+} = \sqrt{  (v_{n+} - v_{s+}) (v_{n+} + v_{s-})}\\
%	N_{e-} = \sqrt{  (v_{n-} - v_{s-}) (v_{n-} + v_{s+}) }\\
%	N_{h+} = \sqrt{  (v_{n+} - v_{s+}) (v_{n+} + v_{s-}) }
%\end{align*}

where, $A = (v_{s-} - v_{n-}) (v_{n+} - v_{s+})$ and $B = (v_{s-} + v_{n+}) (v_{n-} + v_{s+})$. Here, different $v_{n/s \pm}$ denote different velocity components and $N_{e/h \pm}$ are the normalization constants. The explicit expressions of those can be found in Appendix~\ref{AppA}.

%-------------------------------	
\subsection{Lattice model}
%-------------------------------
In order to verify our continuum model based results, we introduce the lattice-regularized version of the Hamiltonian mentioned in Eq.~(\ref{Eq4}). Such lattice Hamiltonian can be written as,

\begin{eqnarray}
	\begin{aligned}
	h^{\tau} &= \lambda_{0} k_{x}^{2}  +   \lambda_{1} \tau  k_{x}^{3} + \tau \Delta_{vp} -\mu \\
			&= -2 \lambda_{0} \text{cos}(k_{x}) + 2 \tau \lambda_{1} \text{sin}(k_{x})\\
			 & - \tau \lambda_{1} \text{sin}(2k_{x}) + \tau \Delta_{vp} - \mu + 2\lambda_{0}\ ,
	\end{aligned}
	\label{Eq9}
\end{eqnarray} 
where, we assume the lattice constant $a = 1$. All the symbols carry their usual meanings as defined before. The real-space version of that Hamiltonian reads as,

\begin{eqnarray}
	\begin{aligned}
		H_{\tau} &=  \sum_{n}\tilde{\mu} c^{\dagger}_{n} c_{n} + \sum_{n} (t c^{\dagger}_{n} c_{n+1} + t^{'} c^{\dagger}_{n} c_{n+2}) + \text{H.c.}\ ,
	\end{aligned}
 \label{Eq10}
\end{eqnarray}

Here, $\tilde{\mu}=\tau \Delta_{vp} - \mu + 2\lambda_{0}$, $t = -\lambda_0 - i\tau \lambda_1 $ and $ t' = \frac{1}{2} i \tau \lambda_1 $ represent the renormalized chemical potential, nearest-neighbor 
and next-nearest-neighbor hopping amplitudes, respectively, in the real-space tight-binding model. We implement this Hamiltonian and calculate the scattering coefficients using the Python package KWANT~\cite{Groth-KWANT}. Finally, we obtain the thermoelectric coefficients by performing numerical integration as discussed in the next section.

%\vspace {-0.55cm}
%-----------------------------------------------

%======================================================
\section{Thermoelectric coefficients }\label{Sec:III}
%======================================================
In this section, we discuss the various thermoelectric coefficients %derived from Onsager relation and 
in terms of the scattering amplitudes based on our setup.

%-------------------------------------------------
\subsection{Calculation of thermal transport}
%-------------------------------------------------
In this subsection, we begin by defining a transmission function for the NS junction due to the transport of electron/hole degrees of freedom across the junction using Blonder-Tinkham-Klapwijk (BTK) formalism~\cite{BTK}. 
This is given by,
 \begin{eqnarray}\label{Eq11}
 	\mathcal{T}(E) = 1 - R_{e}(E) + R_{A}(E)\ , 
 	\label{Eq11}
 \end{eqnarray}
where, normal and AR probability can be defined as, $R_{e}(E)=\left|r_e\right|^2$ and $R_{A}(E)=\left|r_A\right|^2$, respectively. Note that, in the transperant limit (Andreev limit) $R_A = 1$, hence the transmission function becomes $T(E) = 2$. 
In contrast, in the tunneling limit ($R_e = 1$), the transmission function reduces to $T(E) = 0$. Hence, in general, the value of the transmission function lies between $0$ and $2$.

Unlike charge transport, we define thermal transmission function across NS junction due to quasiparticle tunnelling with energy $E>\Delta$ as~\cite{thermal-transmission-coefficient},
  \begin{eqnarray}\label{Eq12}
 	\mathcal{T}_{thermal}(E) = 1 - R_{e}(E) - R_{A}(E) \ .
 	\label{Eq12}
 \end{eqnarray}
 
In the linear response regime, the charge current across a NS junction in the presence of small voltage $\Delta V$ and temperature gradient $\Delta T$ can be written as~\cite{BTK},
\begin{equation}\label{Eq13}
	\begin{aligned}
	I = \frac{2 e}{h} \int_{-\infty}^{\infty} &\Big[f_{N}(E-e \Delta V, T+\Delta T) - f_{S}(E,T) \Big]\\
	& \mathcal{T}(E) dE \ .
\end{aligned}
\end{equation}
 
Similarly, we define the heat current across a NS junction due to a small voltage difference and a temperature difference across the junction as,~\cite{Sivan-Imry},
\begin{equation}\label{Eq14}
\begin{aligned}
	I_{Q} = \frac{2}{h}\int_{-\infty}^{\infty}   &\Big[f_{N}(E-e \Delta V, T+\Delta T) - f_{S}(E,T) \Big]\\
	& (E-\mu_{N}) \mathcal{T}_{thermal}(E) dE \ .
\end{aligned}
\end{equation}

Since the differences in voltage and temperature are assumed to be small, one can employ the Taylor expansion in the Fermi distribution as,
\begin{equation}\label{Eq15}
	\begin{aligned}
		f_{N}(E - e\Delta V, T+\Delta T) &= f_{0}(E,T) - e \Delta V \frac{\partial f_{0}}{\partial E} + \Delta T \frac{\partial f_{0}}{\partial T}\ ,\\
		f_{S}(E,T) &= f_{0}(E,T)\ ,
	\end{aligned}
\end{equation}
where, $f_{0}(E,T)$ denotes the equilibrium Fermi distribution at temperature $T$ and energy $E$. We can rewrite  Eq.~(\ref{Eq13}) and Eq.~(\ref{Eq14}) by incorporating this Taylor expanded
Fermi distribution function as,
\begin{equation}\label{Eq16}
	\begin{aligned}
		I = \frac{2 e}{h} \int_{-\infty}^{\infty} \Big[- e \Delta V \frac{\partial f_{0}}{\partial E} + \Delta T \frac{\partial f_{0}}{\partial T} \Big]
		 \mathcal{T}(E) dE \ ,
	\end{aligned}
\end{equation}
\begin{equation}\label{Eq17}
	\begin{aligned}
		I_{Q} = \frac{2}{h} \int_{-\infty}^{\infty} \Big[- e \Delta V \frac{\partial f_{0}}{\partial E} + \Delta T \frac{\partial f_{0}}{\partial T} \Big](E - \mu_{N})\\
		\mathcal{T}_{thermal}(E) dE \ .
	\end{aligned}
\end{equation}

Comparing Eq.~(\ref{Eq16}) and Eq.~(\ref{Eq17}) using the Onsager relations,
\begin{eqnarray}
	I &=& G \Delta V + L \Delta T\ , \\
	I_Q &=& L_Q \Delta V +  \kappa \Delta T\ ,
\end{eqnarray}

where the symbols $G$, $L$, $L_Q$, and $\kappa$ represent the electrical conductance, Seebeck coefficient, Peltier coefficient, and thermal conductance, respectively, one can obtain the different thermoelectric quantities as,
\begin{eqnarray}
 	G &=& \frac{2 e^{2}}{h} \int_{-\infty}^{\infty} \Big(- \frac{\partial f_{0}}{\partial E} \Big) \mathcal{T}(E) dE\ ,\\
	L &=& -\frac{2 e}{h} \int_{-\infty}^{\infty} \Big(- \frac{\partial f_{0}}{\partial T} \Big) \mathcal{T}(E) dE\ ,\\
   \kappa &=&  \frac{2}{h}\int_{-\infty}^{\infty} (E - \mu_{N})\Big(- \frac{\partial f_{0}}{\partial T} \Big) \mathcal{T}_{thermal}(E) dE\ , \\
 	L_{Q} &=& \frac{2 e}{h} \int_{-\infty}^{\infty} (E - \mu_{N})\Big(- \frac{\partial f_{0}}{\partial E} \Big) \mathcal{T}_{thermal}(E) dE\ .
\end{eqnarray}

Note that, one has to consider separate contributions arising from the two valley degrees of freedom ($\tau = \pm 1$) due to the presence of the term $\Delta_{vp}$ in Eq.~(\ref{Eq4}). Taking this into account, we define the above mentioned coefficients as $G_{tot} = G_{+} + G_{-}$, $\kappa_{tot} = \kappa_{+} + \kappa_{-}$, and $L_{tot} = L_{+} + L_{-}$. For simplicity, we omit the subscript``$tot$” 
in our notation for the remainder of the manuscript.

%-------------------------------------------------------------------------------
\subsection{Calculation of Thermopower, Figure of merit and Lorentz number}
%-------------------------------------------------------------------------------
%Thermopower

The ability of a material to generate an electric voltage difference by maintaining a temperature gradient between the hot and cold junction is called the Seebeck effect, or thermopower. This is a crucial property of thermoelectric materials used to convert heat into electrical energy. We define thermopower ($\mathcal{S}$) as follows,
\begin{eqnarray}
	\mathcal{S} = \frac{L}{G} \ . 
\end{eqnarray}

%~~~~~~~~~~~~~~~~~~~~~~~~~~~~~~~~~~~~~~~~~~~~~~~~~~~~~~~~
%Figure of merit
Another important quantity used to quantify the thermoelectric performance of a device is called the figure of merit, defined as,
\begin{equation}
	zT = \frac{\mathcal{S}^{2} G T}{\kappa}\ .
\end{equation}
Higher values of $zT$ (typically $zT > 1$) indicate better performance, making the device a potential candidate for thermoelectric applications.

In order to explore the deviation of thermoelectric responses (in case of superconducting hybrid junction) from the normal metal, we study the WF law. According to this law, 
the ratio of thermal conductance to electrical conductance is proportional to the absolute temperature, with the proportionality constant known as the Lorentz number. It is defined as,
\begin{equation}
\mathcal{L} = \frac{\kappa}{G T}\ .
\end{equation}
For a normal metal, $\mathcal{L} = \mathcal{L}_{0} = \frac{\pi^{2}}{3} \left( \frac{k_{B}}{e} \right)^{2}$. Deviations from the WF law, specifically when $\mathcal{L}$ falls below $\mathcal{L}_{0}$, 
indicates enhanced thermoelectric performance of the device. However, many materials, such as semiconductors, semi-metallic systems like graphene and Weyl semimetals, and heavy-fermion materials, are known to violate the WF law~\cite{WF-violation1,WF-violation2,WF-violation3}. We also discuss similar violations in the context of MATBG based NS junctions in the next section.

%~~~~~~~~~~~~~~~~~~~~~~~~~~~~~~~~~~~~~~~~~~~~~~~~~~~~~~~~~
%~~~~~~~~~~~~~~~~~~~~~~~~~~~~~~~~~~~~~~~~~~~~~~~~~~~~~~~~~
%======================================================
\section{Numerical Results}\label{Sec:IV}
%======================================================
In this section, we present our numerical results based on calculations performed using both the continuum and lattice models in subsequent manner.
%~~~~~~~~~~~~~~~~~~~~~~~~~~~~~~~~~~~~~~~~~~~~~~~~~~~~~~~~~
%~~~~~~~~~~~~~~~~~~~~~~~~~~~~~~~~~~~~~~~~~~~~~~~~~~~~~~~~~
\subsection{Continuum model }
%-------------------------------------------------
In this subsection, we discuss the results obtained from Eqs.~(\ref{Eq5}) and (\ref{Eq6}) employing the scattering matrix formalism. We calculate various thermoelectric coefficients, including the 
thermal conductance, thermopower, and the figure of merit, among others. Additionally, we examine the deviation of the WF law in our NS junction case compared to the normal metals.

%----------------------------------------------------------------------------------
%----------------------------------------------------------------------------------
\begin{figure}[H]
	\centering
	\subfigure{\includegraphics[width=0.48\textwidth]{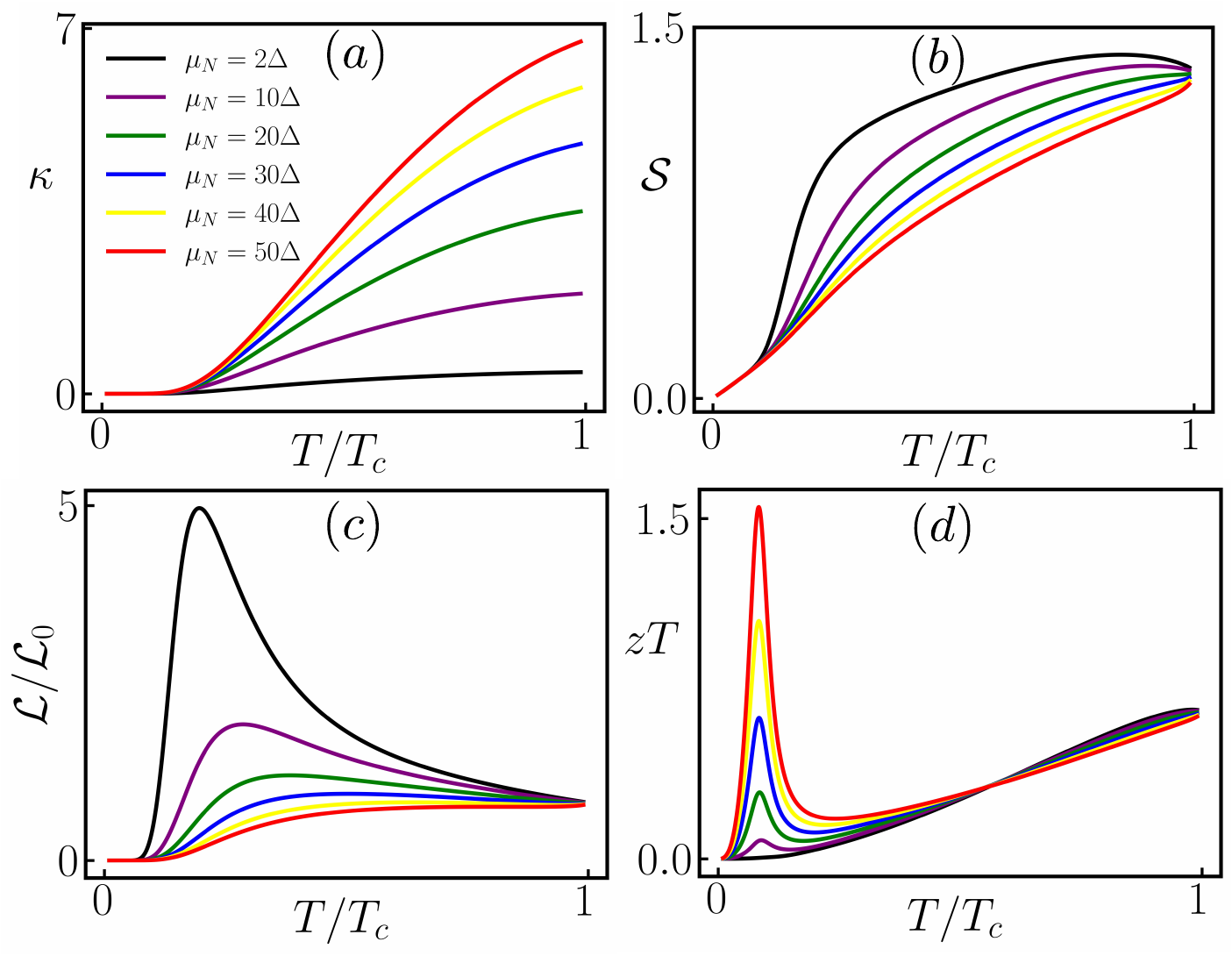}}
	\caption{In Panels (a), (b), (c), and (d), we depict various thermoelectric coefficients: thermal conductance ($\kappa$) in units of $k_{B}^{2}T/h$, thermopower ($\mathcal{S}$) in units of $k_{B}/e$, 
	Lorentz number ($\mathcal{L}/\mathcal{L}_{0}$), and figure of merit ($zT$), respectively, as a function of temperature choosing different values of the chemical potential, using the continuum 
	model analysis. We choose the model parameters as $\lambda_{0} = 125$  in $\Delta \cdot nm^{2}$ unit, $\lambda_{1} = 50$  in $\Delta \cdot nm^{3}$ unit, $\Delta_{vp} = 0$ and $\mu_{s} = 200 \Delta$.
	}
	\label{fig2}
\end{figure}
%------------------------------------------------------------------------------------
%------------------------------------------------------------------------------------
%Thermal conductance from continuum model:
%-------------------------------------------
\subsubsection{Thermal conductance}
%-------------------------------------------
In Fig.~\ref{fig2}(a), we present the thermal conductance ($\kappa$) for the MATBG based NS junction as a function of temperature ($T/T_{c}$) for various values of ($\mu_{N}$) 
tuned in the normal region. For a particular value of $\mu_{N}$, the thermal conductance decreases exponentially with reducing temperature, reflecting the $s$-wave nature of the superconductor.
When $\mu_{N}$ is comparable to $\Delta$, the AR probability $R_{A}(E)\sim 1$ and thermal conductivity becomes vanishingly small as evident from Eq.~(\ref{Eq12}) where $R_{e}(E)\sim 0$. 
However, increasing the chemical potential in the normal region reduces the AR contribution and enhances quasiparticle tunneling. 
%The Andreev reflection reaches its maximum value ($\approx 1$) when the chemical potential in the normal region equals that of the superconducting region, i.e., $\mu_N \approx \mu_s$. 
This is also evident from Eq.~(\ref{Eq8}), which leads to increased electron tunneling into the superconducting region. Therefore, the thermal conductance progressively increases with increasing 
$\mu_N$, as indicated by the transition from the black to red curves. Note that, when $\mu_N \approx \mu_s$, only normal tunneling contributes to thermal conductance $\kappa$.

%Thermopower and zT from continuum model:
%---------------------------------------------------
\subsubsection{Thermopower and Figure of merit}
%---------------------------------------------------
In Fig.~\ref{fig2}(b) and Fig.~\ref{fig2}(d), we depict the thermopower ($\mathcal{S}$) and figure of merit ($zT$) as a function of temperature for various values of $\mu_N$, similar to the thermal conductance. When $T<T_{c}$ (intermediate temperature regime) and $\mu_{N}\sim \Delta$, the transmission function $\mathcal{T}(E)$ is dominated by the AR probability $R_{A}(E)$. Hence, both electrical conductance ($G$) and Seebeck coefficient ($L$) are comparable to each other and their ratio (\ie $\mathcal{S}$) becomes large. As one increases $\mu_{N}$ in the normal region, 
$R_{A}(E)$ becomes small and quasiparticle tunneling starts dominating and the corresponding ratio of $L$ and $G$ turn out to be smaller compared to the previous case.  As $T\sim T_{c}$,                                   $\mathcal{S}$ reaches to a saturated value due to normal tunneling for all $\mu_{N}$.
%Since thermopower is inversely proportional to electrical conductance ($G$), and the amplitude of the electrical conductance increases with increasing $\mu_N$, we observe that thermopower %decreases with increasing $\mu_N$ in the intermediate temperature regime. Hence, near the charge neutrality point (i.e. $\mu_{N} = 0$), we obtain a large thermopower. 
On the other hand, $zT$ becomes larger in the lower temperature regime, decreases towards the intermediate temperature range, and increases again at higher temperatures 
as the chemical potential increases. Notably, around $T = 0.2T_{c}$, $zT$ exceeds unity for $\mu_{N} = 50 \Delta$, while for $\mu_{N} = 10 \Delta$, it reaches $\sim~0.7$ at $T = T_{c}$. 
At lower temperatures, when $\mu_{N}\gg \Delta$, quasiparticle tunneling contribution results in an enhancement in $zT$.

%the temperature-dependent superconducting gap is maximized, which enhances the quasiparticle contribution, resulting in an enhancement in $zT$.
%WF-law from continuum model
%----------------------------------------
\subsubsection{WF law}
%----------------------------------------
In Fig.~\ref{fig2}(c), we showcase the Lorentz number as a function of temperature choosing different values of $\mu_N$. It is evident that near the charge neutrality point (i.e., $\mu_{N} \approx 0$), 
the Lorentz number deviates significantly from that of a metal in the intermediate temperature range, indicating a clear violation of the WF law. However, as we increase $\mu_N$, approaching the 
metallic regime, the WF law is restored as $\mathcal{L}$ approaches $\mathcal{L}_{0}$. In the low-temperature regime, the WF law is violated due to the presence of superconducting correlations. 
As the temperature increases near $T \approx T_c$, the contribution of the superconducting gap vanishes, and the system behaves like a normal metal, thereby satisfying the WF law.

%--------------------------------------------
%--------------------------------------------

%------------------------------------------------
\subsection{Lattice model}
%-------------------------------------------------
In this subsection, we further verify our continuum model results using a lattice regularized version of the continuum Hamiltonian [described in Eq.~(\ref{Eq10})].

%---------------------------------------------------------------------------------------------
%---------------------------------------------------------------------------------------------
\begin{figure}[h]
	\centering
	\subfigure{\includegraphics[width=0.48\textwidth]{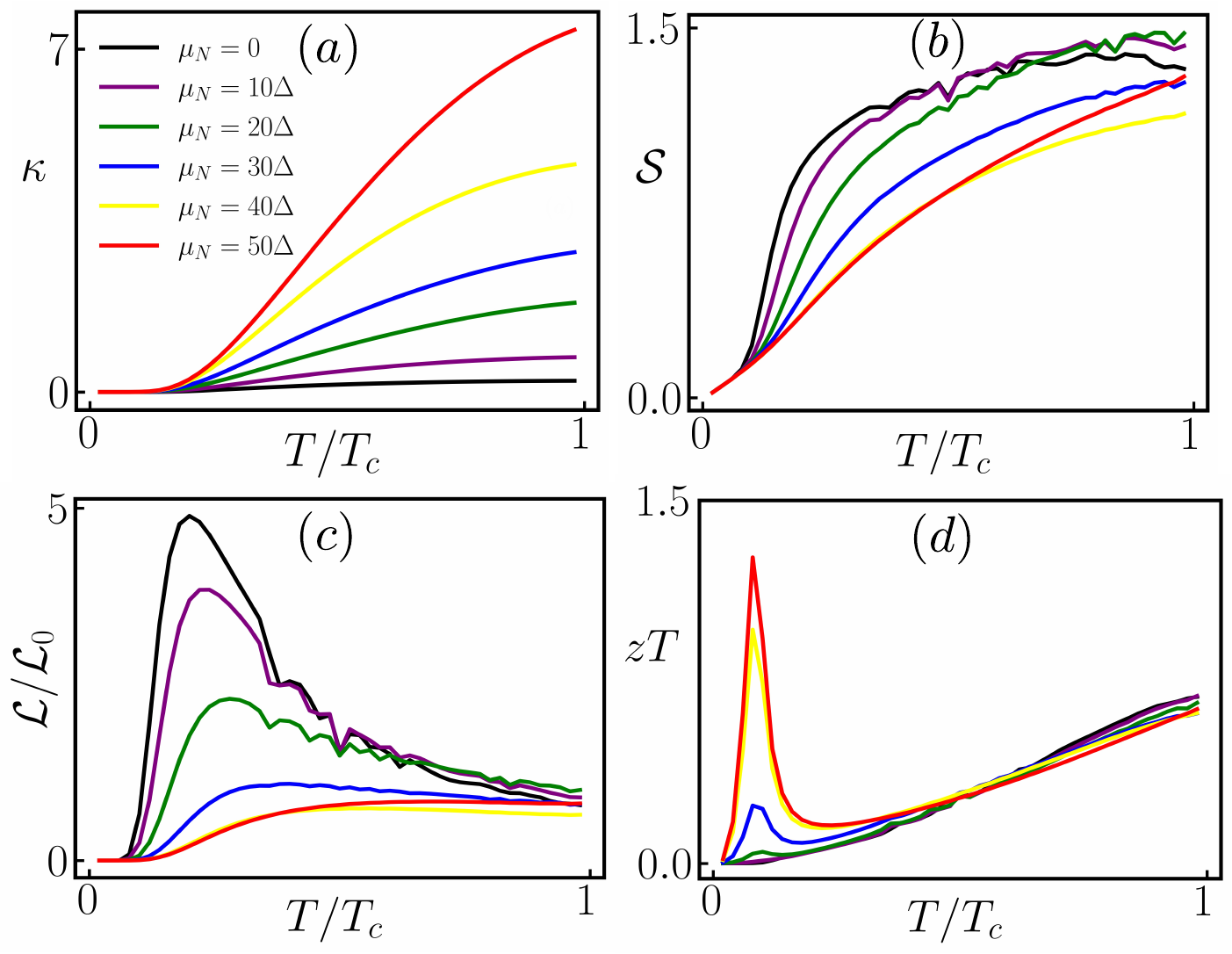}}
	\caption{In panels (a), (b), (c), and (d), we illustrate various thermoelectric coefficients: thermal conductance ($\kappa$) in units of $k_{B}^{2}T/h$, thermopower ($\mathcal{S}$) in units of 
	$k_{B}/e$, Lorentz number ($\mathcal{L}/\mathcal{L}_{0}$), and figure of merit ($zT$), respectively, as a function of temperature for different values of the chemical potential, using the 
	lattice model simulation for a finite system size with \( l = 200 \) lattice sites (measured in units of the lattice constant). We consider the other model parameters to be same as mentioned for the 
	continuum model (see Fig.~\ref{fig2}).}
	\label{fig3}
\end{figure}
%---------------------------------------------------------------------------------------------
%---------------------------------------------------------------------------------------------

%Thermal conductance from lattice model:
%------------------------------------------
\subsubsection{Thermal conductance}
%------------------------------------------
 In this subsection, we discuss the variation of thermal conductance as a function of temperature for different values of $\mu_N$, as depicted in Fig.~\ref{fig3}(a). Note that, for a fixed $\mu_N$ value, 
 the qualitative behavior of $\kappa$ closely resembles the results obtained from continuum model analysis, irrespective of the magnitude. Similarly, in the lattice model, the thermal conductance increases with enhancing $\mu_N$, just as in the continuum model. Therefore, the results obtained from the lattice model align fairly well with the continuum model picture.

%Thermopower and zT from lattice model:
%------------------------------------------------------
\subsubsection{Thermopower and Figure of merit}
%------------------------------------------------------
In Fig.~\ref{fig3}(b) and Fig.~\ref{fig3}(d), we present the thermopower and figure of merit that have been calculated from the lattice model simulation. The thermopower results for different values 
of $\mu_N$ follow a similar trend as the continuum model results, except for some finite-size effects and thermal fluctuations in the high-temperature regime. Other than that, the figure of merit results qualitatively agree with that of the continuum results. In the low-temperature regime, $zT$ increases with increasing $\mu_N$, reaching its maximum value ($zT \approx 1.3$) around $\mu_N = 50\Delta$. In the high-temperature region, $zT$ also increases with $\mu_N$ due to quasiparticle tunneling. However, a closer inspection reveals a quantitative mismatch between the continuum and lattice model results. This discrepancy arises because the effect of the trigonal warping term in the continuum model [i.e. $\lambda_{1}$ in Eq.~(\ref{Eq4})] is renormalized within the band velocity 
[i.e. $v_{\tau, \alpha}$ in Eq.~(\ref{Eq5})], whereas no such approximation is applied in the lattice model. See the latter text for discussion regarding effects due to trigonal warping and valley polarization.
%In the next subsection, we discuss the effects of trigonal warping and valley polarization in greater detail. 

%Wiedemann Franz law
%-----------------------------
\subsubsection{WF law}
%-----------------------------
In this subsection, we again revisit the validity of the WF law based on the lattice model. In Fig.~\ref{fig3}(c), we depict the corresponding Lorentz number as a function of temperature for different 
values of $\mu_N$. Here, we again observe a clear violation of the WF law near the charge neutrality point ($\mu \approx 0$) and in the intermediate temperature regime.  As before, the system approaches $\mathcal{L}/\mathcal{L}_0 \approx 1$ around $T \approx T_c$ and as $\mu_N$ approaches $\mu_S$, indicating that the system approaches the metallic regime.

\vspace{0.3cm}
%~~~~~~~~~~~~~~~~~~~~~~~~~~~~~~~~~~~~~~~~~~~~~~~~~~~~~~~~~~
%~~~~~~~~~~~~~~~~~~~~~~~~~~~~~~~~~~~~~~~~~~~~~~~~~~~~~~~~~~
\begin{figure}[htb!]
%\hspace{-1.2cm}
	\centering
%	\hspace{-2.0cm}
	\subfigure{\includegraphics[width=0.5\textwidth]{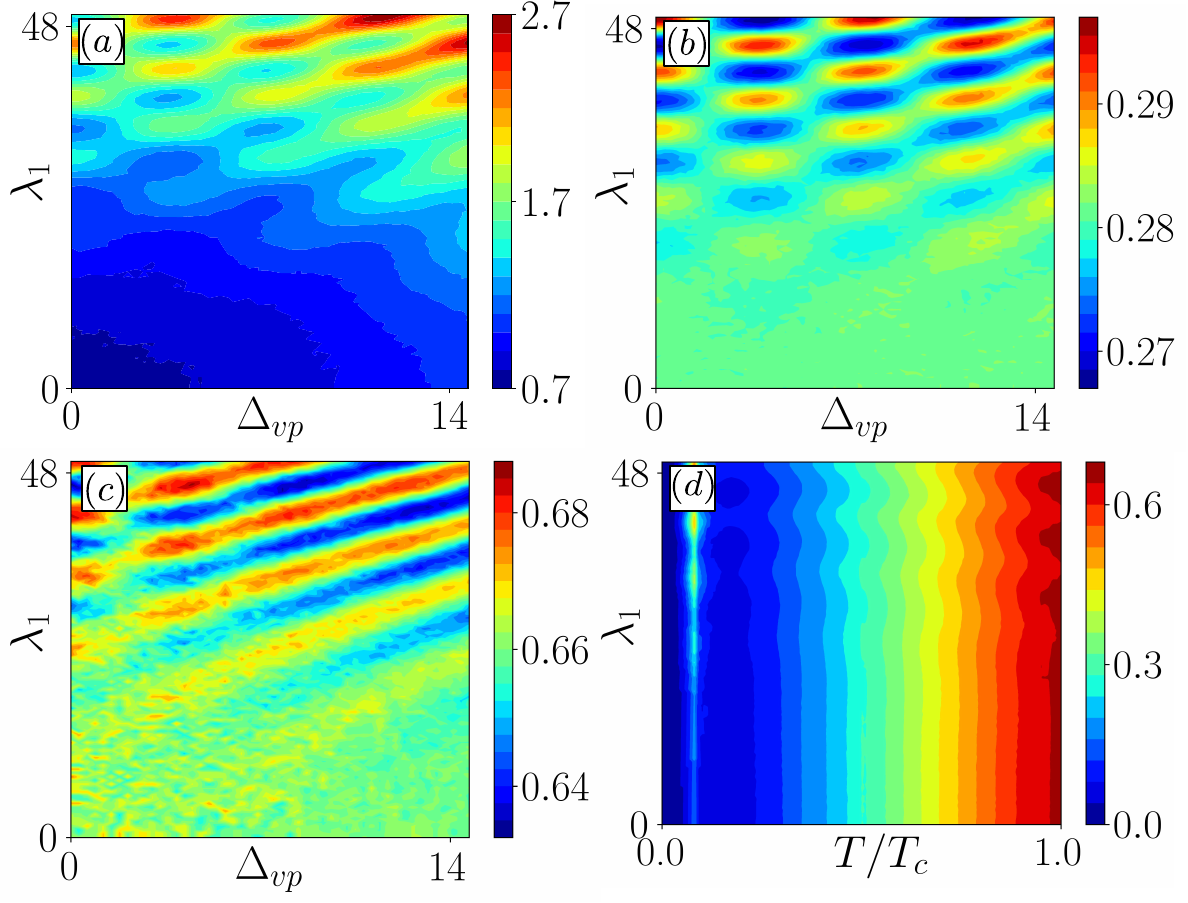}}
	\caption{Panels (a), (b), and (c) demonstrate the density plots for $zT$ in the $\lambda_1 - \Delta_{vp}$ plane choosing three different sets of temperature and $\mu_N$ values: (i) $T = 0.2T_{c}$ 
	and $\mu_{N} = 10 \Delta$, (ii) $T = 0.5T_{c}$ and $\mu_{N} = 30 \Delta$, and (iii) $T = 0.99T_{c}$ and $\mu_{N} = 50 \Delta$, respectively. In contrast, panel (d) exhibits the density plot for 
	$zT$ in the $\lambda_1 - T/T_c$ plane with $\Delta_{vp} = 0$ and $\mu_{N} = 30\Delta$. Here, $\lambda_1$ in $\Delta \cdot nm^{3}$ unit. We consider the lattice model with a finite size of $l = 200$ sites (measured in units of the lattice constant) 	
	%is used for these calculations 
	and other model parameters remain same as mentioned for the continuum model in Fig.~\ref{fig2}.}
 \label{fig4}
\end{figure}
%~~~~~~~~~~~~~~~~~~~~~~~~~~~~~~~~~~~~~~~~~~~~~~~~~~~~~~~~
%~~~~~~~~~~~~~~~~~~~~~~~~~~~~~~~~~~~~~~~~~~~~~~~~~~~~~~~~~

%----------------------------------------------------------------------------
\subsubsection{Effect of trigonal warping and valley-polarization on $zT$}
%----------------------------------------------------------------------------
In this subsection, we discuss the impact of trigonal warping and valley polarization terms on the thermoelectric figure of merit, $zT$, calculated employing the lattice model [Eq.~(\ref{Eq10})]. Here, Figs.~\ref{fig4}(a), \ref{fig4}(b), and \ref{fig4}(c) display density plots for figure of merit $zT$ in the $\lambda_1-\Delta_{vp}$ plane for three different sets of temperature and chemical potential ($\mu_N$) values: (i) 
$T = 0.2 T_{c}$, $\mu_{N} = 10 \Delta$; (ii) $T = 0.5 T_{c}$, $\mu_{N} = 30 \Delta$; and (iii) $T = 0.99 T_{c}$, $\mu_{N} = 50 \Delta$. In each case, we observe several regions with higher $zT$ values (highlighted in red), reaching a maximum of approximately $zT \approx 2.5$ particularly in Fig.~\ref{fig4}(a). These regions emerge as a result of the interplay between trigonal warping ($\lambda_1$) and valley polarization ($\Delta_{vp}$), offering tunability over $ zT $. This indicates that in these regions, $\lambda_1$ and $\Delta_{vp}$ significantly enhance AR over normal reflection, resulting in increased $G/\kappa$ ratio and a high $zT$ value. A brief discussion related to this scenario is provided in Appendix~\ref{AppB}. On the other hand, Fig.~\ref{fig4}(d) depicts the density plot of $zT$ in the $\lambda_1-T/T_{c}$ plane. Here, $zT$ exhibits a monotonic increase with temperature $T/T_c$, except in the low-temperature regime where a discontinuity in the monotonic behavior is observed. Upon closer inspection, such discontinuity in the density plot along the $\lambda_1 = 50$ (in $\Delta \cdot \text{nm}^3$) line in Fig.~\ref{fig4}(d) corresponds to a peak in the line plot depicted in Fig.~\ref{fig3}(d) for $\mu_N = 30\Delta$. The height of this peak increases with the enhancement of chemical potential ($\mu_N$), mainly because, in the low-temperature region, the thermal conductance ($\kappa$) remains very small in magnitude regardless of the chemical potential, while the electrical conductance ($G$) increases with $\mu_N$, resulting in an enhanced $G/\kappa$ ratio. As shown in Fig.~\ref{fig3}(b), in the low temperature regime, thermopower ($S$) almost varies linearly with $T/T_c$. Therefore, $zT$ is primarily influenced by $G/\kappa$ ratio and is enhanced in the low-temperature region, depending on the values of the chemical potential ($\mu_N$) and trigonal warping ($\lambda_1$). In the remaining region of the density plot, $zT$ increases monotonically, reaching its highest values as $T$ approaches $T_c$. This monotonic enhancement of $zT$ can be attributed to quasiparticle tunneling as the temperature approaches $T_c$. 

\vspace{0.2cm}
%-------------------------------------------------------------
%-------------------------------------------------------------
\begin{figure}[H]
	\centering
	\subfigure{\includegraphics[width=0.49\textwidth]{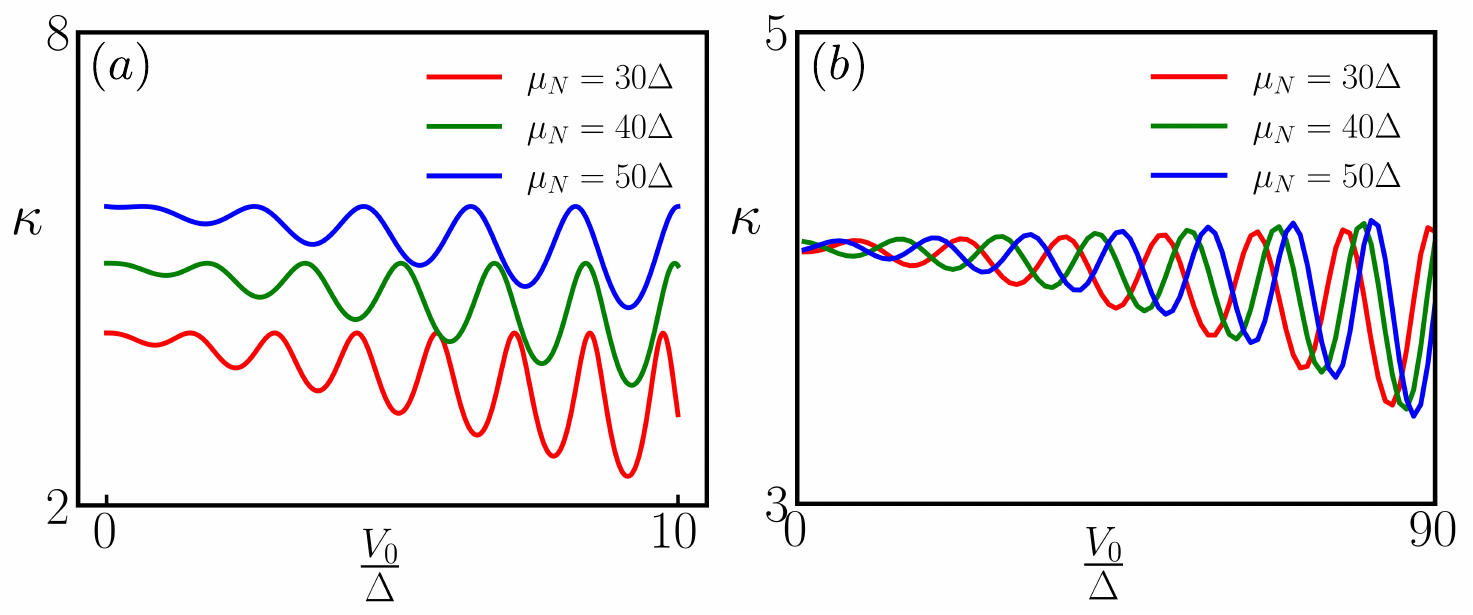}}
	\caption{In panel (a), we illustrate the thermal conductance $\kappa$ (in units of $k_{B}^{2}T/h$) as a function of barrier strength ($V_0$) for different values of $\mu_N$, calculated from the 
	continuum model Hamiltonian. In panel (b), we depict the same quantity calculated employing the lattice model for a system with size $l = 200$ sites (measured in units of the lattice constant) with $T = 0.8T_{c}$ and $d = 10$ nm. 
	For both the continuum and lattice model, the system parameters are chosen to be the same as mentioned in Fig.~\ref{fig2} and Fig.~\ref{fig4}.}
 \label{fig5}
\end{figure}
%--------------------------------------------------------------
%--------------------------------------------------------------

%--------------------------------------------------------
\section{Effect of an insulating barrier}\label{Sec:V}
%--------------------------------------------------------
In this section, we briefly discuss the effect of an insulating barrier (strength $V_0$ and width $d$) inserted at the interface of MATBG based NS junction incorporating both the continuum and lattice models. Here, we assume $\Delta_{vp}=0$. In Fig.~\ref{fig5}(a), we present the behavior of the thermal conductance as a function of barrier strength, calculated from the continuum model Hamiltonian [Eq.~(\ref{Eq5})] for three values of $\mu_N$. Interestingly, the thermal conductance exhibits a damped oscillatory behavior with respect to $V_0$. While such oscillations are commonly seen in massless Dirac fermions with linear dispersion (e.g., graphene or Weyl semimetals), they may seem unexpected in a flat-band system. However, low-energy excitations in MATBG can be described by slow Dirac fermions~\cite{slow-Dirac-Fermions} with a Fermi velocity significantly lower than that of single-layer graphene (approximately 25 times slower). This characteristic makes the oscillatory behavior of the thermal conductance less surprising. If a free electron with energy $E$ encounters a potential barrier of strength $V_{0}$ (where $V_{0} < E$), the transmission probability typically decays exponentially with barrier strength if the wave vector \( k \propto \sqrt{E - V_{0}} \), characteristic of Schr\"{o}dinger-type dispersion. Conversely, an oscillatory transmission probability arises if the dispersion follows \( k \propto (E - V_{0}) \), as in Dirac/Weyl systems~\cite{Bhattacharjee2007,thermal-silicene_GC_PAUL,Weyl-junction-thermal}. In our case, free electrons exhibit an 
\( E \propto k^{2} + k^{3} \) type of band dispersion [as shown in Eq.~(\ref{Eq4})], which encompasses both linear and quadratic band features. Consequently, this hybrid type of dispersion leads to the observed damped oscillatory behavior in the thermal conductance. Also, the magnitude of $\kappa$ increases due to quasiparticle tunneling as $\mu_{N}$ increases further compared to $\Delta$. A phase diagram for the thermal conductance in the plane of barrier potential $V_0$ and barrier width $d$ is provided in Appendix~\ref{AppC}.

%~~~~~~~~~~~~~~~~~~~~~~~~~~~~~~~~~~~~~~~~~~~~~~~~~~~~~~~~~

We further verify our results using the lattice model [Eq.~(\ref{Eq10})], as shown in Fig.~\ref{fig5}(b). Here, the thermal conductance also exhibits a damped oscillation with respect to the barrier 
potential strength, \(V_0\). The results obtained from the lattice model qualitatively match with those of the continuum model, but there appear some quantitative deviations. The possible reason behind such deviation can be attributed to the absence of higher-order terms in the continuum model band structure compared to the lattice model.

%~~~~~~~~~~~~~~~~~~~~~~~~~~~~~~~~~~~~~~~~~~~~~~~~~~~~~~~~~
%\vspace{-0.8cm}
%======================================================
\section{Summary and discussion}\label{Sec:VI}
%======================================================
In this article, we investigate the thermoelectric properties of MATBG based NS hybrid junctions. In the first part of the manuscript, we calculate various thermoelectric coefficients using the scattering matrix approach considering the linearized version of the MATBG. We further verify our results using a lattice-regularized version of the effective continuum model in the latter part 
of the manuscript. Additionally, we discuss the effects of trigonal warping, valley polarization, and the impact of an insulating barrier on the thermal conductance through our hybrid setup. 
%inserted between the NS junction.
Interestingly, we observe a violation of the WF law near the charge neutrality point ($\mu_N \approx 0$); however, for larger $\mu_N$ values, the law holds as one asymptotically approaches the metallic regime. This violation arises because near the charge neutrality point, MATBG electrons behave as slow Dirac fermions. The latter can be further confirmed by the damped oscillatory behavior of the thermal conductance with respect to the barrier strength when an insulating barrier is inserted at the interface of the NS junction (see Fig.~\ref{fig5}). Additionally, we observe a higher $zT$ value due to the interplay between trigonal warping and valley polarization (especially when $T\ll T_{c}$), suggesting that our hybrid setup can serve as a promising thermoelectric candidate with potential device applications. As high thermoelectric figure of merit ($zT$) indicates efficient conversion of heat into electrical energy, enabling MATBG-based NS junctions promising candidates for thermoelectric generators. This opens up the possibilities for designing heat engines and cooling systems using MATBG-based superconducting junctions. In recent years, various devices based on MATBG are fabricated. Specially devices like MATBG based Josephson junctions and SQUIDs are still needed a clear understanding from the thermoelectric aspects for better control and efficient use. 
%of the thermoelectric properties for better control and efficient use. 
Hence our findings may provide valuable insights that can aid in optimizing the performance of such devices. Thermoelectric properties such as thermopower and $zT$ can be effectively tuned by manipulating parameters like valley polarization, trigonal warping, and chemical potential-all of which are influenced by the twist angle. Note that, at other magic angles too, the flat bands are expected to remain isolated from higher-energy bands, and the band symmetries are preserved. Consequently, the effective tight binding model (i.e., Eq.~(\ref{Eq1})) for MATBG on the honeycomb lattice remains unchanged. However this variation in the twist angle leads to renormalized hopping amplitudes. While there is no explicit functional relationship between the twist angle and the parameters of the effective model, it is evident that the values of these parameters-such as $t_1$ , $t_2$ , $\tilde{t_2}$ in the tight-binding model and $\lambda_0$ , $\lambda_1$ in the continuum model should vary 
with the magic angle. This tunability enables the design of customizable thermoelectric devices optimized for specific operating conditions.

%CAR in MATBG:
While in this manuscript we study the thermoelectric properties of a MATBG-based NS junction, based on the local Andreev reflections as the key phenomenon, non-local thermoelectric properties-arising from crossed Andreev reflection (CAR) can play a pivotal role 
in both device applications and in exploring fundamental aspects of non-local quantum transport~\cite{CAR1,CAR2}. It would be interesting to extend our investigations to a MATBG-based Cooper pair beam splitter geometry, where intriguing effects originating from the flat-band physics 
of twisted bilayer graphene could provide deeper insights into entangled electron pair production via non-local CAR processes.

%Experimental possibility
As far as practical feasibility of our hybrid setup is concerned, very recently, several studies have reported experimental investigations of MATBG based hetero junctions~\cite{JJ1_xpt, JJ2_xpt, JJ3_xpt, JJ4_xpt}. Therefore, fabricating MATBG based NS junctions in presence of a gate tunable barrier potential seem feasible. On the other hand, experiments on thermoelectric properties and the thermal diode effect in hybrid junctions~\cite{NIS-junction-thermal1, NIS-junction-thermal2, thermal-diode1, thermal-diode2} have been reported in recent past. Based on this, we believe that it may be possible 
to fabricate our proposed system and conduct thermoelectric measurements to testify our findings. Typically, for a representative superconducting gap value of $\Delta \sim 1$~meV, the figure of merit can reach upto $zT \sim 1.3$ at low temperatures ($T \sim 1$~K) near $\mu_N = 50$~meV and $\mu_S = 200$~meV.

%{\textcolor{red}{{\bf{Mention some realistic numbers from our numerial analysis}}}} For a representative superconducting gap value of $\Delta \sim 1$~meV, the figure of merit reaches $zT \sim 1.3$ at %low temperatures ($T \sim 1$~K) near $\mu_N = 50$~meV and $\mu_S = 200$~meV.

%======================================================
\subsection*{Acknowledgments}
%======================================================
K.B. and A.S. acknowledge the SAMKHYA: High- Performance Computing facility provided by the Institute of Physics, Bhubaneswar and the two Workstations provided by the Institute of Physics, Bhubaneswar from the DAE APEX Project for numerical computations. K.B. and P.C. acknowledge Amartya Pal for stimulating discussion. 

%K.B. and A.S. acknowledge SAMKHYA: High-Performance Computing Facility provided by Institute of Physics, Bhubaneswar, for numerical computations. K.B. and P.C. acknowledge Amartya Pal 
%for useful discussion. 

%======================================================
\subsection*{Data Availability Statement}
%======================================================
The datasets generated and analyzed during the current study are available from the corresponding author upon reasonable request.

%                    Appendix
%%%%%%%%%%%%%%%%%%%%%%%%%%%%%%%%%%%%%%%%%%%%%%%%%%%%%%%%%%%%%%%%%%%%%%%%%%%%%%%%%%%%%
\appendix
%~~~~~~~~~~~~~~~~~~~~~~~~~~~~~~~~~~~~~~~~~~~~~~~~~~~~~~~~~~~~
\section{Scattering States} \label{AppA}
%~~~~~~~~~~~~~~~~~~~~~~~~~~~~~~~~~~~~~~~~~~~~~~~~~~~~~~~~~~~~
As outlined in the main text, the Hamiltonian [Eq.~(\ref{Eq5})] adopts suitable forms in different regions, namely the normal, insulating, and superconducting regions. Below, we explicitly present the 
wave-functions derived and utilized in the scattering matrix calculations discussed in the main text.

The scattering states in the normal region (i.e. $x \leq -d$) are obtained as, 
\begin{eqnarray}
	\psi^{N}_{e, \tau \alpha} = \frac{1}{\sqrt{N_{e, \tau \alpha}}} \left( \begin{array}{c}
		1 \\
		0\\
	\end{array}\right)\ e^{i\alpha k^{N}_{e,\tau \alpha} x}\ ,
	\label{App:1} 
\end{eqnarray}
\begin{eqnarray}
	\psi^{N}_{h, \tau \alpha} = \frac{1}{\sqrt{N_{h, \tau \alpha}}} \left( \begin{array}{c}
		0 \\
		1\\
	\end{array}\right)\ e^{-i\alpha k^{N}_{h,\tau \alpha} x}\ ,
	\label{App:2} 
\end{eqnarray}
The corresponding wave vectors in this region take the following form,
\begin{eqnarray}
	k^{N}_{e, \tau \alpha} = \alpha \frac{\sqrt{\mu_{N} - \tau \Delta_{vp}}}{\sqrt{\lambda_{0}}}+ \frac{\epsilon - \tau \Delta_{vp} }{\alpha v^{N}_{\tau \alpha}}\ , \\
	k^{N}_{h, \tau \alpha} = \alpha \frac{\sqrt{\mu_{N} - \tau \Delta_{vp}}}{\sqrt{\lambda_{0}}}- \frac{\epsilon - \tau \Delta_{vp} }{\alpha v^{N}_{\tau \alpha}}\ .
\end{eqnarray}

Here, $\alpha = \pm 1$ indicates the right- and left-moving states, respectively, while $\tau = \pm 1$ represents the valley degrees of freedom. The corresponding Fermi velocity (longitudinal) 
is given by,
\begin{eqnarray}
	v^{N}_{\tau,\alpha} = -2\sqrt{\lambda_{0} (\mu_{N} - \tau \Delta_{vp}) } + \frac{3\alpha(\mu_{N} -\tau \Delta_{vp} )}{\lambda_{0}} \tau \lambda_{1}\ ,
\end{eqnarray}
where, $\mu_{N}$, $\lambda_{0}$, and $\lambda_{1}$ denote the chemical potential in the normal region, kinetic energy and the trigonal warping term in the bare Hamiltonian, respectively. 
The expression for the normalization constants are given by, 
\begin{eqnarray}
N_{e,\tau \alpha}=N_{h,\tau \alpha} = \sqrt{  (v^{N}_{\tau,\alpha} - v^{s}_{\tau,\alpha}) (v^{N}_{\tau,\alpha} + v^{s}_{\tau,-\alpha})}\ .
\end{eqnarray}

In presence of the proximity induced superconducting term, the wave-functions for the superconducting region (i.e. $x \geq 0$) can be obtained as,

\begin{eqnarray}
	\psi^{s}_{qe, \tau \alpha} = \left( \begin{array}{c}
		e^{-i \alpha \beta}\\
		1\\
	\end{array}\right)\ e^{i\alpha k^{s}_{qe, \tau \alpha} x}\ ,
	\label{App:3} 
\end{eqnarray}
\begin{eqnarray}
	\psi^{s}_{qh, \tau \alpha} = \left( \begin{array}{c}
		e^{i \alpha \beta} \\
		1\\
	\end{array}\right)\ e^{-i\alpha k^{s}_{qh, \tau \alpha} x}\ ,
	\label{App:4} 
\end{eqnarray}
where, $\beta$ is defined as,
\begin{eqnarray}
	\beta &= \cos^{-1}(\frac{\epsilon}{\Delta_{s}}) \hspace{5pt}\text{if}~(\epsilon < \Delta_{s})\ ,\\
	&= \cosh^{-1}(\frac{\epsilon}{\Delta_{s}}) \hspace{5pt} \text{if}~(\epsilon > \Delta_{s})\ .
\end{eqnarray}

The corresponding wave vectors in this region take the form, 
\begin{eqnarray}
	k^{s}_{qe, \tau \alpha} = \alpha \frac{\sqrt{\mu_{s}}}{\sqrt{\lambda_{0}}}+ \frac{\sqrt{\epsilon^{2} - \Delta_{s}^{2} }}{\alpha v^{s}_{\tau\alpha}}\ , \\
	k^{s}_{qh, \tau \alpha} = \alpha \frac{\sqrt{\mu_{s}}}{\sqrt{\lambda_{0}}} - \frac{\sqrt{\epsilon^{2} - \Delta_{s}^{2} }}{\alpha v^{s}_{\tau \alpha}}\ .
\end{eqnarray}
where, $\mu_{s}$ is the chemical potential in the superconducting region. Here, ``$qe$'' and ``$qh$'' in the subscript represent respectively the electron like and hole like quasiparticles.

\vspace{0.2cm}
%~~~~~~~~~~~~~~~~~~~~~~~~~~~~~~~~~~~~~~~~~~~~~~~~~
%~~~~~~~~~~~~~~~~~~~~~~~~~~~~~~~~~~~~~~~~~~~~~~~~~
\begin{figure}[H]
	\centering
	\subfigure{\includegraphics[width=0.49\textwidth]{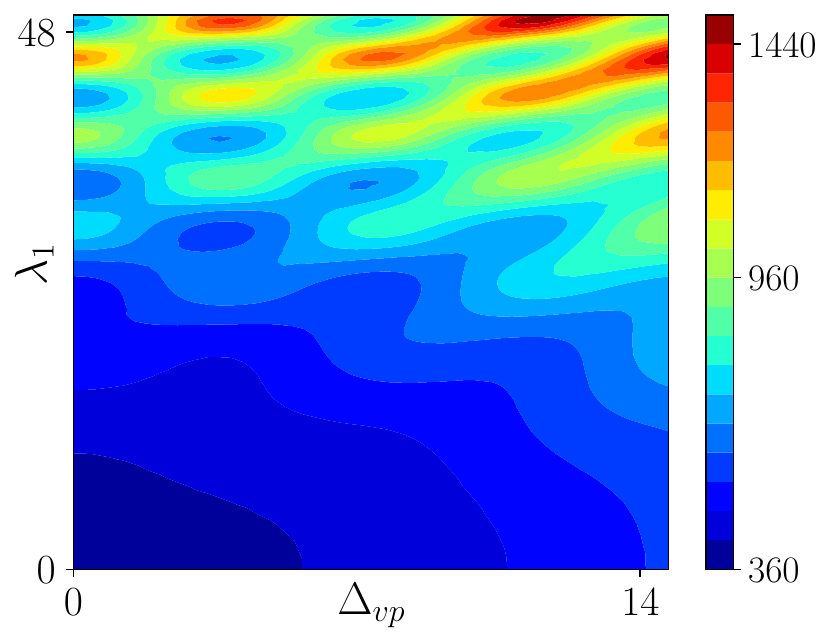}}
	\caption{We illustrate the ratio of electrical conductance $G$ (in units of $e^{2}/h$) and thermal conductance $\kappa$ (in units of $k_{B}^{2}T/h$) \ie ($G/\kappa$) in the plane of trigonal warping 
	($\lambda_{1}$) and valley polarization ($\Delta_{vp}$) choosing chemical potential $\mu_N = 10\Delta$ and $T = 0.2 T_{c}$. We use the same lattice model %for this phase plot keeping the model 
	parameters as mentioned in Fig.~\ref{fig4} of the main text.}
	\label{fig6}
\end{figure}
%~~~~~~~~~~~~~~~~~~~~~~~~~~~~~~~~~~~~~~~~~~~~~~~~
%~~~~~~~~~~~~~~~~~~~~~~~~~~~~~~~~~~~~~~~~~~~~~~~~

To incorporate the effect of the insulating barrier, the chemical potential in the normal region ($\mu_N$) is replaced by $(\mu_N - V_0)$, where $V_0$ denotes the barrier strength. This substitution 
alters the wave vector and, in turn, the scattering states. The wave-functions within the insulating region ($-d \leq x \leq 0$) are then given by,
\begin{eqnarray}
	\psi^{I}_{e, \tau \alpha} = \frac{1}{\sqrt{N_{e, \tau \alpha}}} \left(\begin{array}{c}
		1 \\
		0\\
	\end{array}\right)\ e^{i\alpha k^{I}_{e,\tau \alpha} x}\ ,
	\label{App:1} 
\end{eqnarray}
\begin{eqnarray}
	\psi^{I}_{h, \tau \alpha} = \frac{1}{\sqrt{N_{h, \tau \alpha}}} \left(\begin{array}{c}
		0 \\
		1\\
	\end{array}\right)\ e^{-i\alpha k^{I}_{h,\tau \alpha} x}\ ,
	\label{App:2} 
\end{eqnarray}
Here, the above wave vectors are given by, 
\begin{eqnarray}
	k^{I}_{e, \tau \alpha} = \alpha \frac{\sqrt{\mu_{N} - V_{0} - \tau \Delta_{vp}}}{\sqrt{\lambda_{0}}}+ \frac{\epsilon - \tau \Delta_{vp} }{\alpha v^{I}_{\tau \alpha}}\ , \\
	k^{I}_{h, \tau \alpha} = \alpha \frac{\sqrt{\mu_{N} - V_{0} - \tau \Delta_{vp}}}{\sqrt{\lambda_{0}}}- \frac{\epsilon - \tau \Delta_{vp} }{\alpha v^{I}_{\tau \alpha}}\ .
\end{eqnarray}
Also the corresponding longitudinal Fermi velocity can be written as,
%\textcolor{blue}{
\begin{eqnarray}
	\begin{aligned}
	v^{I}_{\tau,\alpha} = -2\sqrt{\lambda_{0} (\mu_{N} - V_{0} - \tau \Delta_{vp}) } + \\ \frac{3\alpha(\mu_{N} - V_{0} -\tau \Delta_{vp} )}{\lambda_{0}} \tau \lambda_{1}\ .
	\end{aligned}
	\label{EqA18}
\end{eqnarray}
%}
The calculation of different scattering amplitudes for the continuum model Hamiltonian in the main text are performed based on these wave-functions, applying the appropriate boundary condition.

\vspace{0.1cm}
%-----------------------------------------------------------
%-----------------------------------------------------------

%----------------------------------------------------------
%----------------------------------------------------------

%~~~~~~~~~~~~~~~~~~~~~~~~~~~~~~~~~~~~~~~~~~~~~~~~~~~~~~~~~~
\section{Insights into the behavior of $zT$} \label{AppB}
%~~~~~~~~~~~~~~~~~~~~~~~~~~~~~~~~~~~~~~~~~~~~~~~~~~~~~~~~~~
In Sec.~\ref{Sec:IV}B of the main text, we discuss the effects of trigonal warping and valley polarization on the figure of merit ($zT$) of the MATBG based NS junction. To further understand the behavior of $zT$ with varying trigonal warping ($\lambda_1$) and valley polarization ($\Delta_{vp}$), we analyze the ratio of electrical conductance ($G$) to thermal conductance ($\kappa$). Recall the formula $zT = (\mathcal{S}^2 G T)/\kappa$, where \(\mathcal{S}\) is the thermopower and $T$ is temperature of the system. In Fig.~\ref{fig6}, we present a density plot of $G/\kappa$ in the $\lambda_1$-$\Delta_{vp}$ plane, calculated using the same parameter set as described in the main text (i.e. $\mu_N = 10\Delta$ and $T = 0.2 T_{c}$ ). We observe that the density plot for $G/\kappa$ closely resembles the $zT$ results shown in Fig.~\ref{fig4}(a) of the main text. To gather a microscopic understanding of how the varying trigonal warping ($\lambda_1$) and valley polarization ($\Delta_{vp}$) and their interplay impact the AR process and improves the $G/\kappa$, below we provide a detailed discussion.
	
\begin{figure*}[t]
	\subfigure{\includegraphics[width=0.9\textwidth]{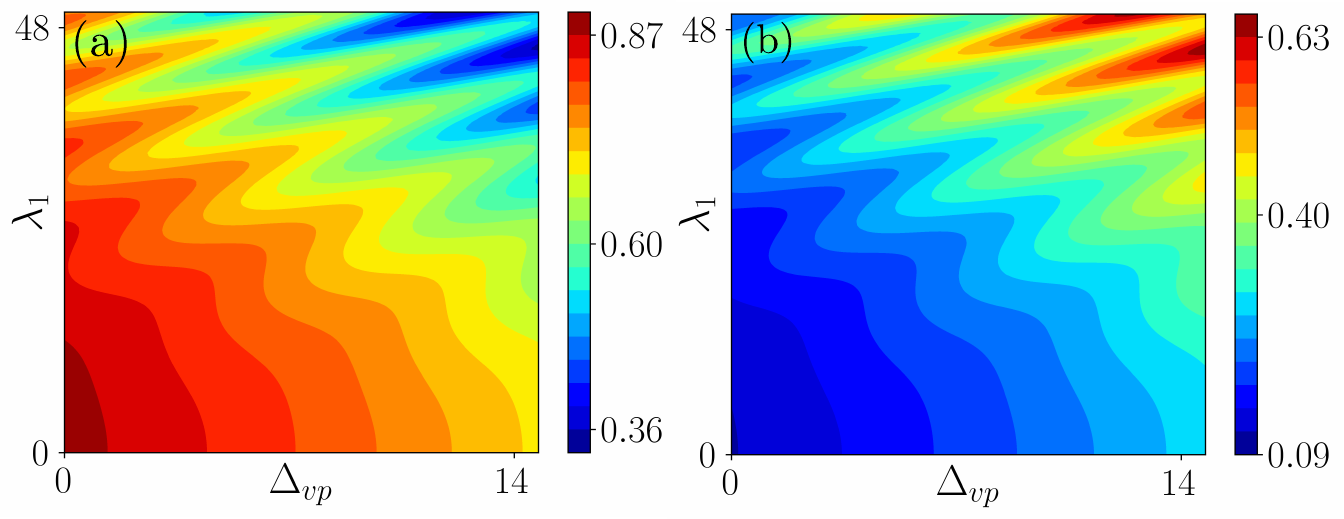}}
	\caption{In panels (a) and (b), we display the probability of normal electron reflection and AR respectively in the $\lambda_1$-$\Delta_{vp}$ plane for a fixed energy $E = 0.5\Delta$ and chemical potential $\mu_{N} = 10 \Delta$.}
	\label{fig7}
\end{figure*}

In Figs.~\ref{fig7} (a) and (b), we respectively display the normal and AR probabilities in the plane of valley polarization ($\Delta_{vp}$) and trigonal warping ($\lambda_1$), at a fixed energy $E = 0.5\Delta$ and chemical potential $\mu = 10\Delta$. To understand the effect of valley polarization ($\Delta_{vp}$), we examine the line for $\lambda_1 = 0$. We observe that, as $\Delta_{vp}$ increases, the normal reflection probability decreases while the AR increases. However, the magnitude of AR probability remains smaller than the probability of normal reflection along this line.

Similarly, by following the line for $\Delta_{vp} = 0$, we explore the effect of trigonal warping. We find that, with increasing $\lambda_1$, the normal reflection probability decreases very slowly, accompanied by a gradual increase in the AR probability, along with an oscillatory behavior. It is also noted that the AR probability is much smaller in magnitude than that of the normal reflection along the $\lambda_1$ axis.

While individual values of $\lambda_1$ and $\Delta_{vp}$ do not induce a dominant AR across the phase diagram, there are regions where normal reflection is suppressed, and AR becomes dominant. This arises due to the interplay between valley polarization ($\Delta_{vp}$) and trigonal warping ($\lambda_1$). These regions correspond to where the $G/\kappa$ ratio improves, thereby enhancing the figure of merit $zT$.

%This indicates that the ratio of electrical conductance ($G$) to thermal conductance ($\kappa$) plays a pivotal role in determining the behavior of the $zT$.
%-----------------------------------------------------------------------------------
%-----------------------------------------------------------------------------------
%----------------------------------------------------------------------------------
%----------------------------------------------------------------------------------

%~~~~~~~~~~~~~~~~~~~~~~~~~~~~~~~~~~~~~~~~~~~~~~~~~~~~~~~~~~~~~~~~~~~~~~~~~~~
\section{Impact of barrier strength and thickness on $\kappa$} \label{AppC}
%~~~~~~~~~~~~~~~~~~~~~~~~~~~~~~~~~~~~~~~~~~~~~~~~~~~~~~~~~~~~~~~~~~~~~~~~~~~

%~~~~~~~~~~~~~~~~~~~~~~~~~~~~~~~~~~~~~~~~~~~~~~~~~~~~~~~~~~~~~~~~~~~~~~~~
%~~~~~~~~~~~~~~~~~~~~~~~~~~~~~~~~~~~~~~~~~~~~~~~~~~~~~~~~~~~~~~~~~~~~~~~~
\begin{figure}[H]
	\centering
	\subfigure{\includegraphics[width=0.49\textwidth]{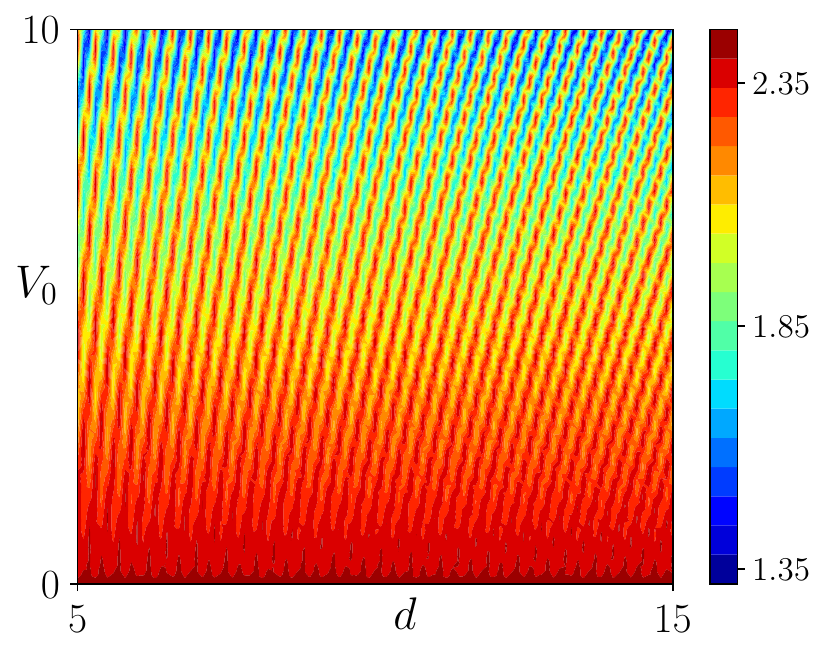}}
	\caption{We depict the thermal conductance $\kappa$ (in units of $k_{B}^{2}T/h$) in the plane of barrier strength $V_0$ (expressed in terms of $\Delta$) and barrier width $d$ (in unit of $nm$) 
		considering chemical potential $\mu_N = 30\Delta$ and $T = 0.8 T_{c}$. We utilize the continuum model for generating this density plot maintaining the same system parameters as mentioned 
		in Fig.~\ref{fig2} of the main text.}
	\label{fig8}
\end{figure}
%~~~~~~~~~~~~~~~~~~~~~~~~~~~~~~~~~~~~~~~~~~~~~~~~~~~~~~~~~~~~~~~~~~~~~~~~~
%~~~~~~~~~~~~~~~~~~~~~~~~~~~~~~~~~~~~~~~~~~~~~~~~~~~~~~~~~~~~~~~~~~~~~~~~~

To further illustrate the impact of the insulating barrier on the MATBG based NS junction, we present a density plot of the thermal conductance ($\kappa$) in Fig.~\ref{fig8} considering the plane of barrier strength ($V_0$) and the barrier width ($d$). In this continuum model based analysis, the chemical potential of the normal region is fixed at $\mu_N = 30\Delta$, and the temperature is set to $T = 0.8T_c$. Here, Fig.~\ref{fig8} exhibits oscillatory behavior in the thermal conductance with the variation of both $V_0$ and $d$. However, the magnitude of $\kappa$ decreases as one increases both $V_0$ and $d$. The reason can be attributed from Eq.~(\ref{EqA18}) where the first term can become imaginary as we increase $V_0$, while the second term remain real giving rise to oscillatory behavior with decaying magnitude of $\kappa$.

%for both scenarios: varying the barrier strength ($V_0$) while keeping the barrier width ($d$) constant, and varying the barrier width ($d$) while keeping the barrier strength ($V_0$) fixed. 
%This oscillatory pattern highlights the intricate dependence of thermal transport on the barrier parameters.

%======================================================

\bibliography{bibfile}{}

%============End of MAIN PAPER=============

\end{document}